\RequirePackage{fix-cm}
\documentclass[smallextended, natbib]{svjour3}       
\usepackage[utf8]{inputenc}
\usepackage{amsmath}
\usepackage{amssymb}
\usepackage{textcomp}
\usepackage{graphicx}
\usepackage{listings}
\usepackage[british]{babel}
\usepackage[left=3cm,right=3cm,top=1cm,bottom=1cm,includeheadfoot]{geometry}
\usepackage[normalem]{ulem} 
\usepackage{color}
\usepackage{listings}

\usepackage{versions}
\includeversion{arXiv}
\excludeversion{journal}

\begin{journal}
  \usepackage[backend=bibtex8,
  style=authoryear-ibid,
  url=false,
  maxnames=99,
  maxcitenames=2,
  firstinits=true,
  dashed=false]{biblatex}
  \AtEveryBibitem{
    \clearfield{month}%
  }
  \DeclareFieldFormat[article,inbook,incollection,inproceedings,patent,thesis,unpublished]{citetitle}{#1\isdot}
  \DeclareFieldFormat[article,inbook,incollection,inproceedings,patent,thesis,unpublished]{title}{#1\isdot} 
  \renewbibmacro{in:}{}
  \DeclareFieldFormat{journaltitle}{#1\isdot}
  \usepackage{csquotes}

\end{journal}

\begin{arXiv}
\end{arXiv}

\newcommand{\mus}{\,\text{\textmu s}}

\renewcommand{\u}[1]{\,\text{#1}}
\newcommand{\mum}{\,\text{\textmu m}}
\newcommand{\tw}{\textwidth}

\newcommand{\blender}{\texttt{blender}}
\newcommand{\stl}{\texttt{stl} geometry}
\newcommand{\ior}{index of refraction}

\begin{document}
\title{
Realistic prediction and analysis of cavitation bubble dynamics via optical ray-tracing of the corresponding numerical simulations }

\titlerunning{Cavitation bubble ray-tracing}   

\author{Max Koch \and
        Juan M. Rossell\'o \and
        Christiane Lechner \and 
        Werner Lauterborn \and
        Julian Eisener \and
        Robert Mettin
}

\institute{Max Koch, Christiane Lechner \at
                Georg-August Universität Göttingen \\
                Drittes Physikalisches Institut\\
                37077 Göttingen -- Germany \\
                Tel.: +49-(0)551-39-27744\\
                \email{max.koch@phys.uni-goettingen.de}
            \and
            Juan M. Rossell\'o, Julian Eisener \at
            Otto-von-Guericke-Universität Magdeburg\\
            Institut für Physik (IfP)\\
            39106 Magdeburg -- Germany
            \and Werner Lauterborn, Robert Mettin \at
            Georg-August Universität Göttingen \\
            Drittes Physikalisches Institut\\
            37077 Göttingen -- Germany \\
            Tel.: +49-(0)551-39-26933\\
            \email{robert.mettin@phys.uni-goettingen.de}
}

\date{Received: date / Accepted: date}

\maketitle

\begin{abstract}
Experimental analysis of cavitation bubble dynamics typically uses optical imaging. However, images are often severely affected by distortions and shadows due to refraction and total reflection at the liquid--gas interface. Here it is shown that optical ray-tracing can be a powerful tool for the analysis process by assisting in the comparison of experiments to numerical two-phase flow simulations.
To this end, realistic ray-tracing images are generated from numerical simulations that correspond to the experimental setup. We apply the method on the jetting dynamics of single bubbles collapsing at a solid wall. Here, ray-tracing can help in the interpretation of raw experimental data, but also in prediction of the complex bubble interface deformations and internal structures during the collapse. The precise shape of the highly dynamical bubbles can be inferred and thereby a correction method for velocity values of the liquid jets can be provided.
\end{abstract}

\section{Introduction and motivation}
Designing advanced experiments needs substantial planning and also knowledge of what results to expect. Testing different configurations can be costly and time consuming, and methods to help in this respect may be welcome. When simulations of two-phase flows and experiments with imaging cameras are involved, a special blend of the visualization of numerical results and the images from the experiment may substantially improve the outcome by taking into account the omnipresent refraction at phase boundaries. Vice versa, the experimental observations could be optimized by analyzing optical ray paths in the arrangement and by later inserting numerical simulations into the ray-tracing engine used. Then the experimental photograph can directly be compared with the simulated image. The abstract concept will become more clear by an example: a bubble in water. 
 
\begin{figure}[htb]
\centerline{\includegraphics[width=0.6\textwidth]{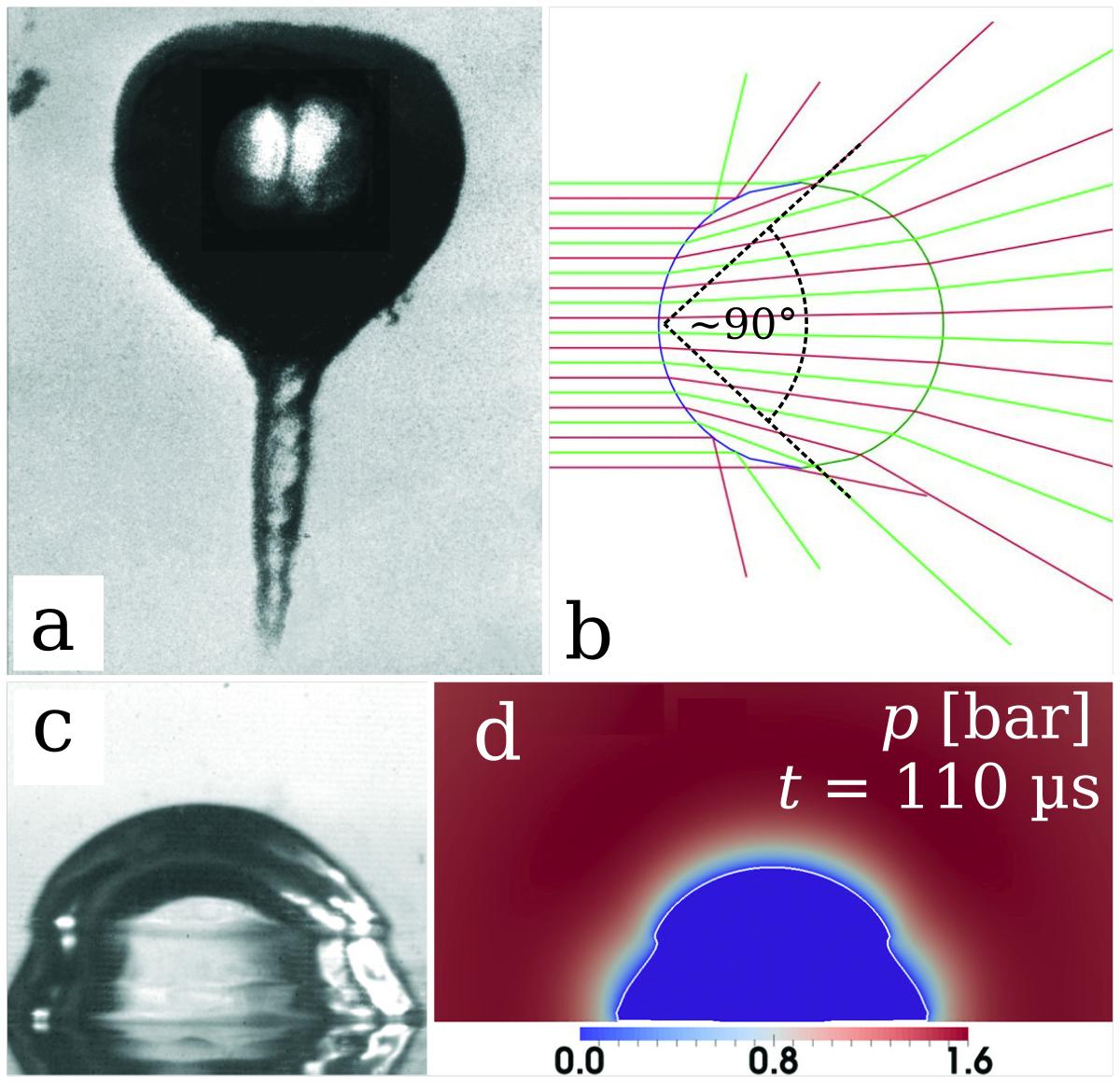}}
\caption{\label{fig:Fig1} 
a) Bubble with jet photographed in backlight (shadowgraphy).
b) Applying Snell's law to a spherical air bubble in water for a bundle of parallel light rays.
c,d) Standard comparison of an experimental, photographic image of a collapsing bubble in water (c) and a color coded visualization of the pressure in a corresponding numerical 
simulation as a cut through the bubble (d). 
}
\end{figure}

\paragraph{The bubble dynamics challenge}
Bubbles in liquids, a huge topic in science and engineering \citep{tgleighton,brennen,lbGreatReview}, are experimentally demanding objects owing to their fast dynamics on small spatial scales with complicated two-phase three-dimensional topology (see, e.g., \citet{Lindau-2003}). Of special interest is the formation of fast liquid jets that appear in non spherically-symmetric environments \citep{Supponen-2016,Rossello-2018}. These jets are one of the reasons for the erosion of hardest materials by bubbles. 
The jets are formed in either of two ways: either by involution of the bubble interface
\citep{PC71,Lauterborn-1975} due to restricted flow by a nearby object, or by impact of an annular inflow that then squeezes out the liquid in the respective orthogonal direction 
\citep{Voinov-1976,Lechner-2019}. After formation, the jet traverses the interior of the bubble and impacts onto the opposite bubble wall. To follow this phenomenon it is necessary to have optical access to the interior of the bubble. Up to now, this is mostly done by diffuse illumination from the back (shadowgraphy) or multiple light sources to better show the topography of the deformed bubble surface \citep{Lauterborn-Goettingen-1980,Reuter-2016}. 
In shadowgraphy, smooth bubbles then appear black on a bright background with a bright center, where the light can pass undeflected (e.g., \citet{Reuter-2016,Rossello-2018}). The dark view of the rest of the bubble interior is due to the light being deflected off the surface of the bubble and not being able to reach the photographic film or the CCD chip of an electronic camera. An iconic, historic example of a bubble jet \citep{Lauterborn-Goettingen-1980} is presented in Fig.~\ref{fig:Fig1}a. The bubble was produced by focused laser light in a process called optical breakdown. The bubble is in its re-expansion phase (rebound) after the first collapse. The jet is due to the asymmetry introduced by a solid boundary below the bubble in the direction of the jet. As only part of the interior of the bubble is optically accessible, the motion of the jet through the bubble can not be followed in its entirety. By special illumination and observation, the very formation of the jet by involution of the top of the bubble could be photographed \citep{Lindau-2003}. 

Regarding cavitation bubble dynamics, most publications focus either on numerical or on experimental results, and very few are known that compare experimental and numerical results directly \citep{Lauterborn-1975, Ohl-2000, Blake-2015, Han-2015}. 
This might be due to historical reasons, but with nowadays technological advancement it became possible to simulate the whole experiment. 

Pursuing the goal to illuminate the whole volume of a small, air-filled sphere in water, a simple ray-tracing calculation based on Snell's law \citep{python-raytracer} would give the graphic in Fig.\,\ref{fig:Fig1}\,b. Here, a case is considered, where the real experiment makes use of a long-distance microscope observing a tiny bubble. 
Thus it can be assumed that only parallel rays will enter an objective to the left of the bubble in the graphic.
It is seen that the objective is hit by rays coming from a conical region behind (to the right) of the bubble. The cone has an aperture angle of about 90$^\circ$. Therefore, in order to illuminate most of the bubble interior an illumination from behind the spherical bubble with the size of the base of that cone would be necessary.
 Still, the interior would be mapped towards the center region in the image seen by the microscope, leaving a thick dark outline at the bubble rim. However, this dark rim of the bubble might again be lightened by oblique incidence of light rays due to total reflection, such that the real interface position may become blurred or unclear (compare top and lowermost three rays of Fig.\,\ref{fig:Fig1}\,b). 

As a general rule, when the shape of the objects under analysis becomes increasingly complex, it is almost impossible to reconstruct the object by simple observation. Fig.~\ref{fig:Fig1}\,c and Fig.~\ref{fig:Fig1}\,d are supposed to show the same bubble generated in direct vicinity to a solid surface, but it is evident that the background illumination is distorted strongly by refraction in the experimental image due to the interface of the bubble.

For many applications a clear view into the interior of the object of interest is required and more sophisticated illumination schemes must be implemented. In this study, we will explore the feasibility of the ray-tracing method to analyse complex scenarios. Specifically, the case of study will be given by a water jet produced when a bubble collapses close to a solid boundary. The scope is to deduce information from the interior of the bubble in order to find a more precise means for jet speed measurements. 

Analytical investigations of Snell's law for a circular interface have been used for a droplet \citep{Kobel-2009} and also for cylindrical or spherical glass vessels filled with sulfuric acid \citep{rossello-2016}. In the latter, this method was combined with stereo camera recording and iterative triangulation to determine the position of a bubble inside the vessel.
It is important to note that the previous analytic methods are useful only in cases with very simple geometries, and thus cannot be applied to a strongly aspherical bubble.

The combination of the ray-tracing method and computational fluid dynamics is found in other works \citep{semlitsch-2010,craig-2016} only in the sense of simulating heat radiation and the reaction of the fluid towards that radiation. 

Recently, the ray-tracing method was used in the same sense as presented here, but for single-phase fluid dynamics: A Schlieren image from Large Eddy simulations was compared with  experiments of a supersonic gas flow by \citet{Luthman-2019}. The authors implemented a ray-tracing engine themselves and found very good agreement of the resulting images between experiment and simulation for several flow features, e.g., jet--shock interaction and inferred velocity fields.

Ray-tracing of the two-phase flow of a single cavitation bubble was briefly introduced twenty years ago by \citet{Ohl-2000}. Applying nowadays computational tools allow for investigations in much higher detail. Here we use \blender, a free and open-source 3D animation tool with a python scripting interface, but the same technique can be also implemented by any optical ray-tracing software. Apart from the enhanced readability of scientific results, it can serve as an alternative validation method of any newly developed (two-phase) computational fluid dynamics simulation code by comparison with the experiment. Another possibility is to redesign experiments in order to optimize the visualization of specific expected phenomena. 

This work is organized as follows: The experimental setup and the ray-tracing method are described in Sec.~\ref{sec:exp-setup}. The bubble model for the simulations and the numerical implementation are introduced in Sec.~\ref{sec:code}. In the results sections, the method is demonstrated in three steps: 
First, the ray-tracing engine is validated via an experiment by using a static bubble and varying the position of part of the illumination (Sec.~\ref{sec:results_static}). Second, with the confidence in the ray-tracing engine an expanding and collapsing bubble is investigated, which during collapse produces a pronounced liquid jet. This bubble will be referred to as the \emph{jet-illusion bubble} (Sec.~\ref{sec:jet-illusion-bubble}) because a distorted jet image leads to an \emph{illusional} high jet velocity impression.
Here, the method helps to infer the bubble shape to an unprecedented extent in precision from the experiment and thereby correct the velocity value of the jet speed.
In a third step, the overlay-method is applied to an even more difficult case (Sec.~\ref{sec:fast-jet-bubble}), a second type of expanding and collapsing bubble, in the following termed the \emph{fast-jet bubble}, which upon collapse is supposed to exhibit an extensively fast, but hardly measurable jet \citep{Lechner-2019}. Both the ray-traced numerical simulation and the experimental results match very well, suggesting that the fast jet predicted by the numerical simulation also exists in the real experiment.

\section{Experimental setup and ray-tracing method}\label{sec:exp-setup}
The overlay-method is demonstrated by two different experiments. In the first experiment, a static bubble is investigated for two different locations of a second illumination flash. Using a static bubble avoids complicated time dependencies and makes optical studies easier. The bubble is generated with a micro-liter syringe inside a cubical glass cuvette (Hellma) filled with de-ionized and degassed water. In the second experiment, a dynamical bubble with strong expansion and collapse is investigated. It is generated by optical breakdown in water, induced by a focused nanosecond pulse of a frequency-doubled Nd:YAG laser (Litron Nano PIV) at 532\u{nm} wavelength with a pulse duration of $\approx 10\u{ns}$. In both experiments, an Imacon 468 high-speed camera is used with a \emph{K2 infinity} long-distance microscope attached, including a 4x magnifying objective allowing for a resolution of 2\mum\ per pixel. Eight pictures can be taken at constant pixel resolution ($385\times 575\u{px}$). The maximum time resolution used per measurement was 150\,ns for the exposure time per image and 350\,ns between end and start of two successive pictures, resulting in 2\,Mfps.

\subsection{Experimental setup for the static bubble}
\label{sec:staticbubble}
\begin{figure}[htb]
     \begin{center}
            \includegraphics[width=0.7\tw]{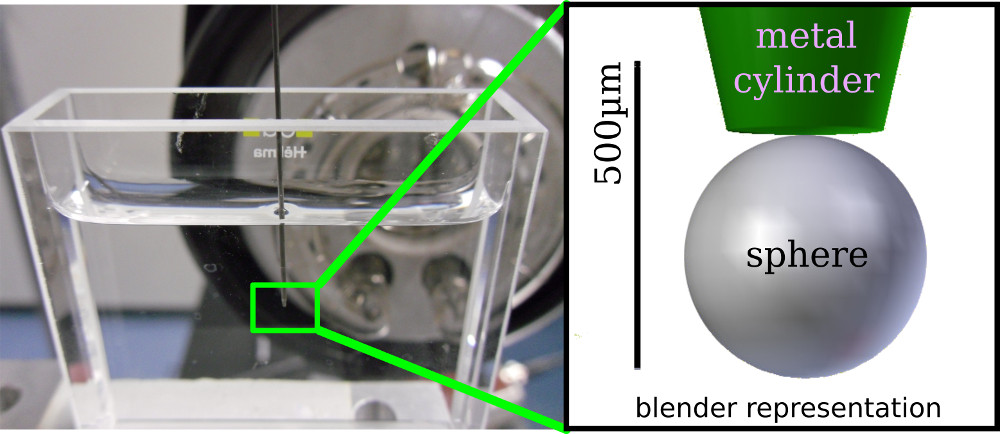}
     \caption{ Comparison of the experimental arrangement (photograph), left, and corresponding numerical configuration for \blender, right, of a static, spherical bubble to validate the \blender\ ray-tracing engine. 
     }
     \label{fig:setup_static_bubble_photo}
     \end{center}
\end{figure}

In order to validate the \blender\ engine, an experiment was chosen,
where a bubble of less than a millimeter in diameter rests fixed on a microliter syringe needle (Fig.\,\ref{fig:setup_static_bubble_photo}). This static bubble then has a very low contact area to the needle and is therefore almost perfectly spherical in shape. This scene can be mimicked in the \blender\ software as shown in Fig.\,\ref{fig:setup_static_bubble_photo}, right.

\begin{figure}[htb]
    \begin{center}
        \includegraphics[width=0.35\tw]{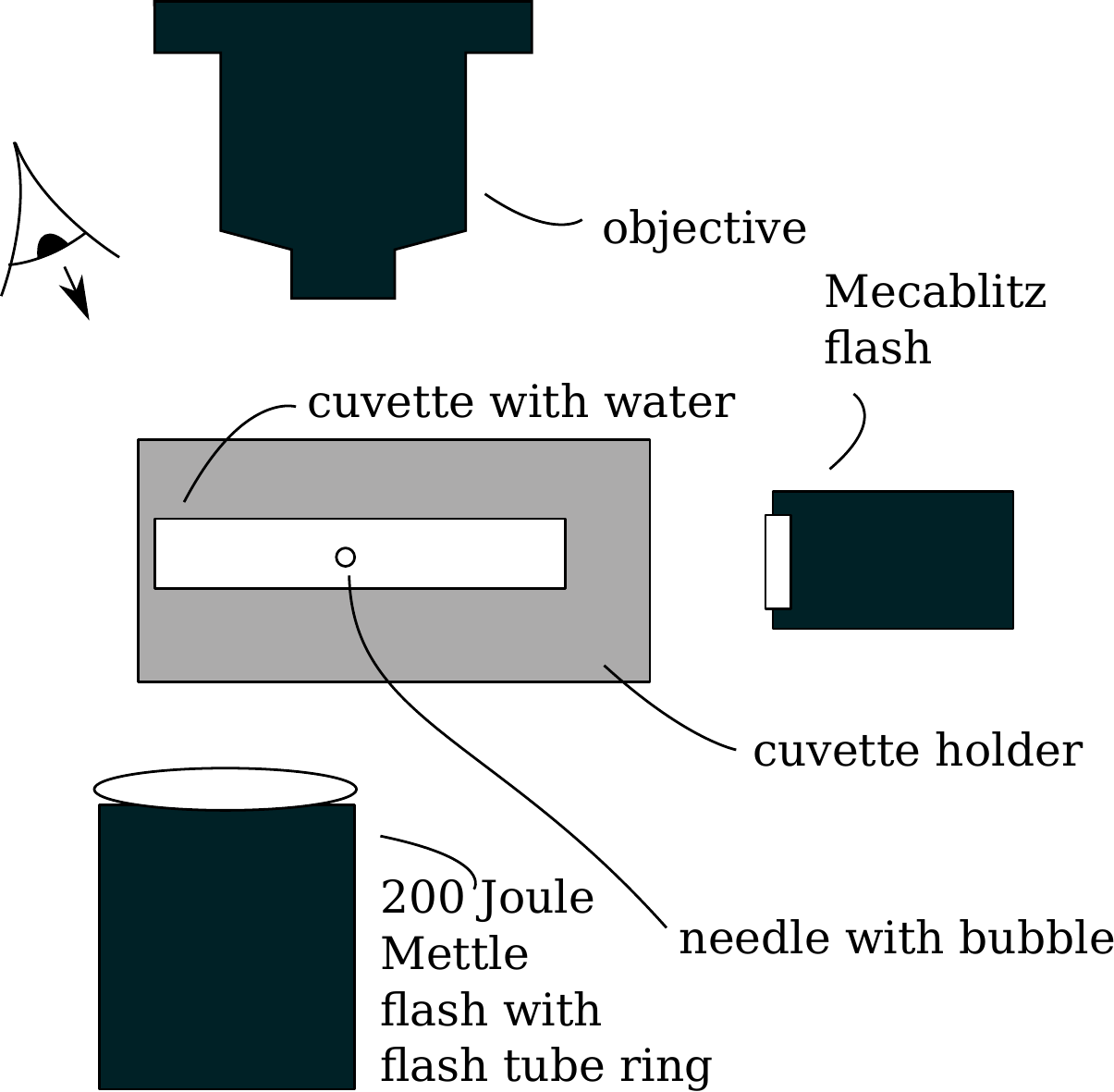}
        \includegraphics[width=0.34\tw]{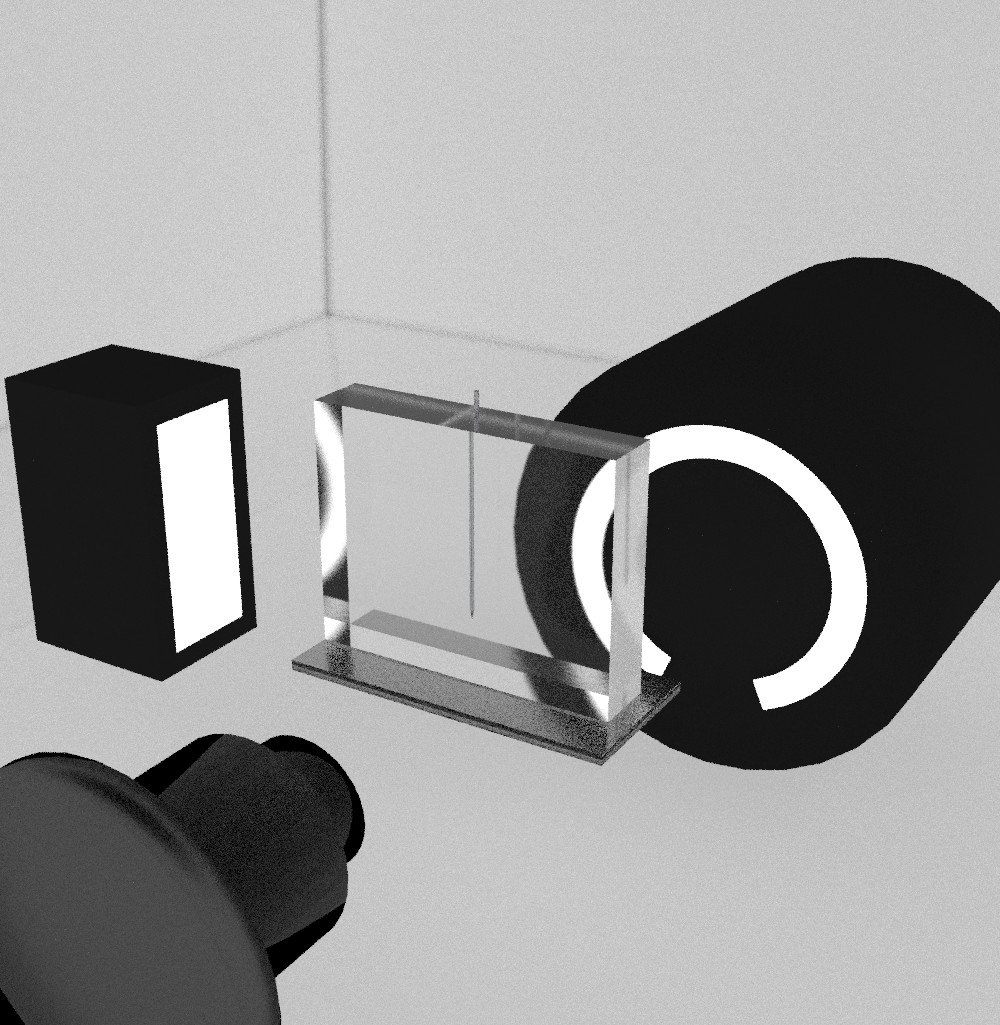}
    \end{center}
    \caption{Setup for the static bubble experiment. Sketch of the elements of the experimental arrangement (left) transferred into \blender\ elements (right) from the perspective indicated by the eye in the sketch. Needle and water block are modeled as well in \blender.}
    \label{fig:setup_static_bubble}
\end{figure}

The complete experimental setup (see Fig.\,\ref{fig:setup_static_bubble}) consists of essentially six elements:
(i) The water filled glass cuvette with inner size of $1\u{cm}\times 5\u{cm} \times 4\u{cm}$ in length, width and height, and wall thickness of 2\,mm; 
(ii) a background illumination xenon flash \textit{Mettle MT-600DR} emitting about 200\u J of light over about 8\u{ms} by a ring flash tube. The distance to the bubble was 5.2\u{cm}.
(iii) A side illumination xenon photo flash (\textit{Mecablitz 36CT2}) with a straight flash tube of 35\,mm length emitting $\approx 40\u J$ and Fresnel lens (distance to the bubble $\approx 8.5\u{cm}$); 
(iv) a microliter syringe and needle producing a bubble of about 300\u{\textmu m} to 500\u{\textmu m} in diameter inside the cuvette; 
(v) a K2 Infinity microscope objective as described above; 
(vi) and a high speed camera (\textit{Imacon 468}) as described above.

\subsection{Experimental setup for the two dynamical bubbles}
\begin{figure}[ht]
    \begin{center}
         a)   \includegraphics[width=0.38\tw]{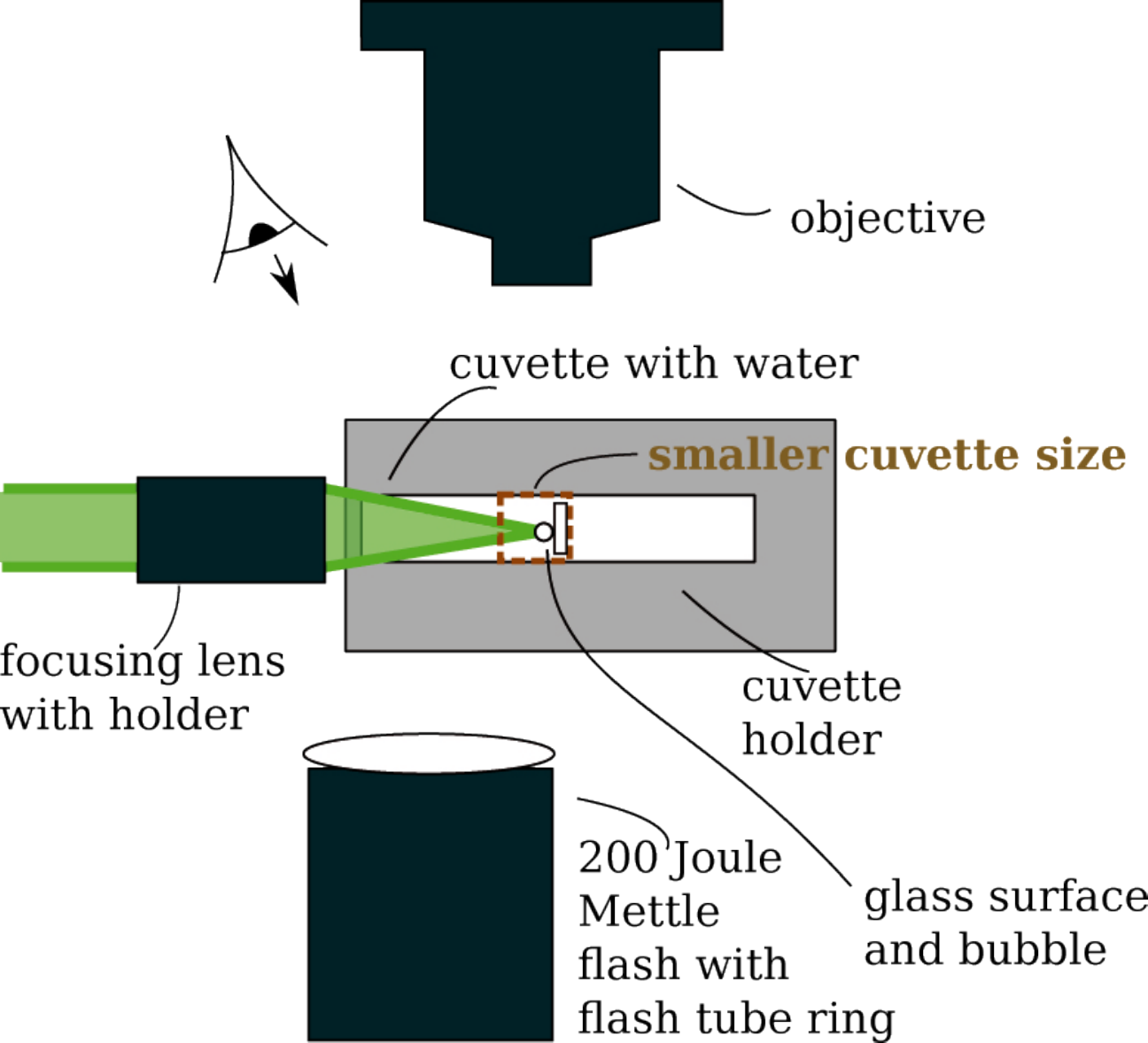}
         b)   \includegraphics[width=0.3\tw]{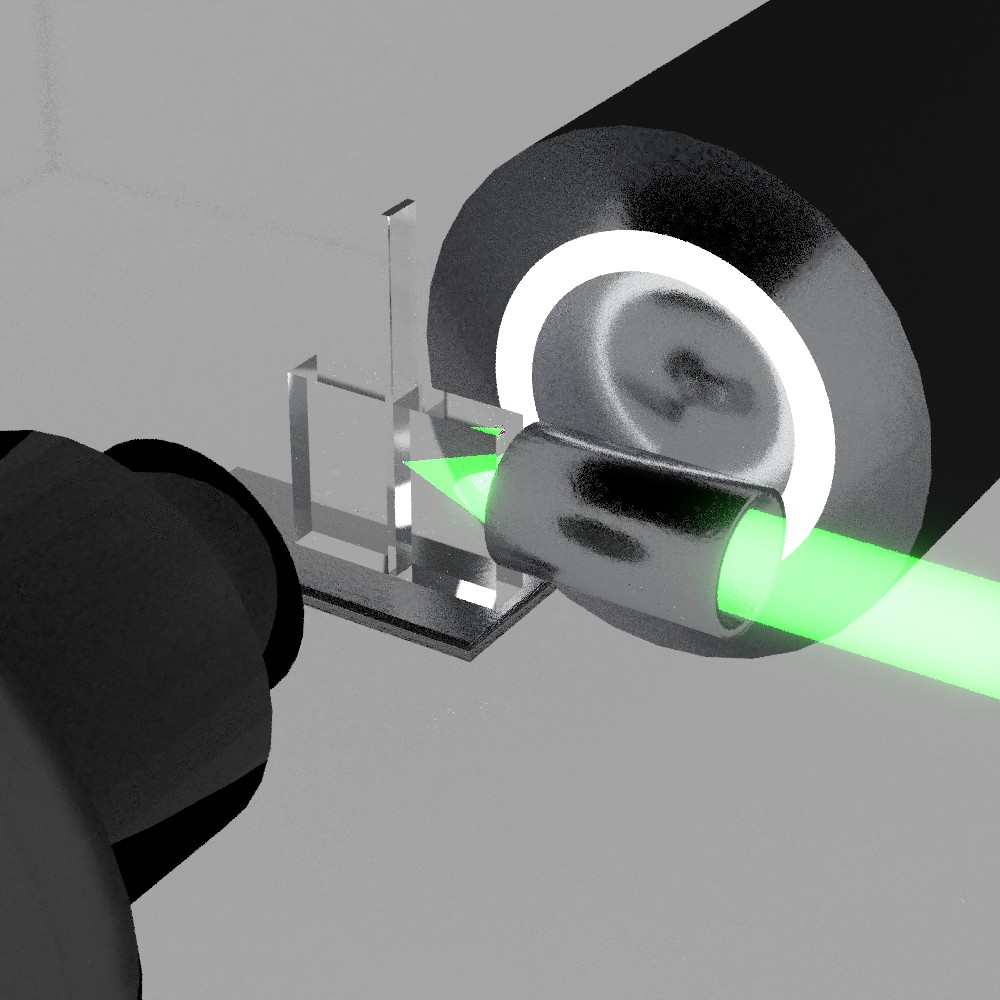}
         c)   \includegraphics[width=0.22\tw]{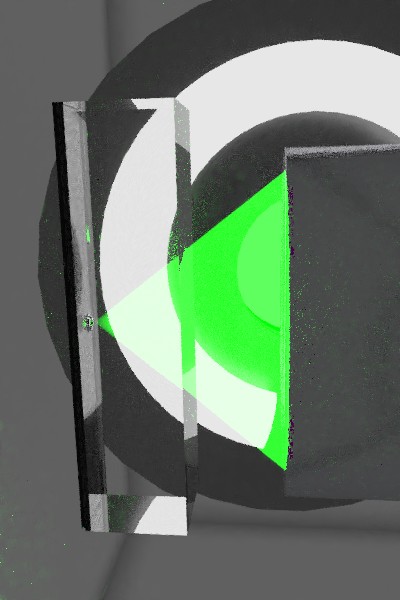}
    \end{center}
    \caption{Sketch (a) and \blender-ray-traced representation (b),(c) of the experimental setup producing Figs.~\ref{fig:THE_JET} and \ref{fig:thejetfusion}. The dynamical bubble is generated with a nanosecond pulse of a frequency doubled Nd:YAG laser with pulse energy of about 30\,mJ. The bubble is produced on a glass surface put vertically into the cuvette. Illumination is done by the Mettle flash with ring xenon tube.
    }
    \label{fig:setup_dynamical_bubble}
\end{figure}

Having validated the \blender-ray-tracing engine for a static, sub-millimeter bubble (see Section~\ref{sec:results_static}), two experiments of highly dynamical cavitation bubbles are performed, using ultra-high-speed (mega frames per second) photography. The setup for the two experiments is the same, shown in Fig.~\ref{fig:setup_dynamical_bubble}, except for the size of the cuvette. 
The setup resembles the one in Fig.\,\ref{fig:setup_static_bubble}, but the side flash, as well as the syringe needle are now omitted.
The first experiment (with smaller cuvette size $1\u{cm}\times 1\u{cm} \times 4\u{cm}$ in length, width and height) gives insight into a liquid jet which seems to be at least twice as fast due to optical distortions, the second experiment (cuvette size as for the static bubble experiment) shows the detailed dynamics of a bubble on a solid surface. 
In both setups, the single cavitation bubble is now generated by the Nd:YAG laser at an approximate energy of 30\,mJ. The light pulse is focused into the cuvette in order to induce a plasma spot by optical breakdown \citep{Lauterborn-1974,lbvogelshockwavebook,kkdiss}. 
 
The light from the Nd:YAG laser is focused by a lens with a short focal length directly onto the surface of a planar glass object, located at the rear side of the smaller cuvette of the first experiment and in the middle of the cuvette in the second experiment. 
The position of laser produced a bubble that is expanding and collapsing directly at the wall corresponds to a dimensionless wall distance of \begin{align}
    d/R_{max}=D^\ast. \label{eq:dstar}
\end{align}
Here $d$ is the distance of the laser plasma to the wall, and $R_{max}$ is the theoretical maximum bubble radius before collapse in unbounded liquid. The resulting nearly hemispherical bubble dynamics with complicated collapse dynamics and potential development of a fast, thin jet has been described in
\citep{Lechner-2019}. 
A general overview on the dependence of bubble collapse dynamics on $D^\ast$ is given in 
\citep{LechnerArXiv-2020}.

\subsection{Overlay-method recipe for the ray-tracing}
The idea is to generate an image from the results of a (two-phase) computational fluid dynamics simulation that looks most alike the one obtained from the experiment in order to bring both to an overlay. Assuming that the geometry and dimensions of the experiment including the illumination devices are known, the procedure for the numerical side would then be the following:

Firstly, the closed interface iso-plane of the two fluid phases from the numerical CFD simulation has to be extracted to a standard 3D format. For example, most programs are suited with an import/export function to the \texttt{stl} format. 
Secondly, the \texttt{stl} file obtained can be imported into a program with a realistic, light-ray-tracing engine.  A variety of specialized ray-tracing programs might exist. 
Here, the free, open-source project \blender\ is taken because it incorporates a very realistic and physical lighting engine, standard \texttt{stl} import compatibility, a python-language application programming interface (API) and a large variety of 3D editing tools. The latter become important when it comes to modelling the experimental setup, and the API is handy for batch operations on many simulation time steps and parameter scans.
Note that the aim is to produce a realistic image from the CFD simulations rather than perform scientific analyses of optics. Thus interference can be neglected, but the intensity and diffusivity of the refraction and reflection at objects matter. 
The following steps were performed with \blender\ in order to achieve a realistic image:

The \texttt{cycles} render engine is used \citep{cycles-render-engine}. This engine emits the light rays from the camera into the 3D scene and distinguishes between so called camera rays, reflected rays, transmission rays and shadow rays. 
After importing the \stl\ object its surface is both smoothed and reduced in complexity by applying the \textit{limited dissolve} algorithm. This algorithm accounts for reducing the amount of faces while keeping the same shape. This \stl\ is then given a material with an index of refraction (IOR). The so called \textit{GlassBSDF} material with an \ior\ of $0.75$ suits best for an air bubble in water 
Depending on the direction of the face normal of the edited \stl\ object, the IOR ratio either has to be set to 1.333 or $1/1.333=0.75$.

The optically relevant geometries of the experiment then are created around the bubble geometry along with their optical properties of, e.g., glossyness, light transmission or light emission. Simple diffusive, glossy, glass-like or emissive materials do the work in most cases to mimic optically relevant lab equipment. The water of the cuvette is mimicked by a block given the \textit{GlassBSDF} material with an \ior\ of $1.333$. This block can be seen in Figs.~\ref{fig:setup_static_bubble} and \ref{fig:setup_dynamical_bubble}. The solid boundary made of glass, where the bubble collapses to in the experiment, is modelled by a simple block object of the \textit{GlassBSDF} material with an \ior\ with respect to water of $1.333/1.45=0.92$.
The flash tube geometries can be designed adequately and a simple \textit{emissive} material can be attached to them. The ray-tracing camera can be set up essentially with the same properties as in the experiment concerning focal length, sensor size and pixel resolution.

\section{Bubble simulation code\label{sec:code}}
The Navier-Stokes equation and continuity equation of the compressible, two-phase flow of a single, laser-generated cavitation bubble are solved with the finite volume method (FVM) together with the volume of fluid (VoF) approach \citep{Weller-2008} within the OpenFOAM framework \citep{Weller-1998}. The code has been extensively tested. A brief description will be given here. The reader is referred to \citet{Koch-2016,Lechner-2017, Lauterborn-2018,  Lechner-2019} for a more detailed description and application.

The asymmetric dynamics of a bubble close to a solid boundary is of special interest due to the liquid-jetting phenomenon. The liquid is compressible for inclusion of pressure waves up to weak shock waves. Surface tension and gravity can be included. The equations are formulated for a \emph{single} fluid satisfying the continuity, Navier-Stokes and material equations
\begin{align}
  \frac{\partial \rho}{\partial t}  +  \nabla \cdot (\rho \, \vec U)  = 0 \, ,  \qquad 
  \frac{\partial (\rho \, \vec U)}{\partial t} + \nabla \cdot (\rho \, \vec U \otimes \vec U)  
  = - \nabla p + \rho \, \vec g + \nabla \cdot  \mathbb{T} + \vec f_{\sigma}, \label{eq:contNSEq}       \\
    \rho_g(p)  = \rho_{n}{\left(\displaystyle \frac{p}{p_n} \right)^{1/\gamma_g}} ,\qquad
    \rho_l(p) = \rho_\infty\left(\frac{p+B}{p_\infty+B}\right)^{1/n_\mathrm T}, \label{eq:tait}
\end{align}
with $\rho, p, \vec U$ denoting the density, pressure and velocity,
$\nabla$ denotes the gradient, $\nabla \cdot$ is the divergence and
$\otimes$ the tensorial product. 
$\mathbb{T}$ is the viscous stress tensor of a Newtonian fluid, 
and $\vec f_{\sigma}$ denotes the force density
due to surface tension and $\vec g$ the gravitational acceleration.
The pressure and the density of the gas in the bubble at normal conditions are denoted by $p_n$ and $\rho_{n}$, respectively, and $\gamma_g = 1.4$ denotes the ratio of the specific heats of the gas (air). 
For the liquid, the Tait equation of state (Eq.~\ref{eq:tait}, right) for water is used (see, e.g., \citet{Fujikawa-1980}) with $p_\infty$ the atmospheric pressure, $\rho_\infty$ the equilibrium
density, the Tait exponent $n_{\mathrm T} = 7.15$ and the Tait pressure $B = 305\,$MPa. 
A volume fraction field $\alpha_l$ with $\alpha_l = 1$ in the liquid phase
and $\alpha_l= 0$ in the gas phase is introduced to distinguish between the
two phases. Density and viscosity fields then can be written as 
$\rho = \alpha_l \rho_l + (1-\alpha_l) \rho_g$, 
$\mu= \alpha_l \mu_l + (1-\alpha_l) \mu_g$. A transport equation for
$\alpha_l$ is obtained from the continuity equation for the liquid phase,
which reads $\partial_t (\alpha_l \rho_l)   +   \nabla \cdot (\alpha_l \rho_l \vec U)   = 0$ when mass transfer across the interface is neglected.
A pressure-based formulation of the equations is discretized with the FVM in the \texttt{foam-extend} fork \citep{fext40} of OpenFOAM.

In order to capture the dynamics of millimeter sized, laser-generated bubbles in water it is sufficient to assume isentropic conditions (heat conduction and viscous dissipation can be neglected, the shock waves in the liquid are of weak strength with only a small increase in entropy across the shock). Therefore, barotropic equations of state are provided for the liquid and the gas and the energy equation is dropped. With this model, excellent agreement with experimental data is found for bubbles close to solid boundaries \citep{Koch-2016} concerning the bubble shape as a function of time, up to jet impact.

The simulations of the first bubble in the results Section~\ref{sec:jet-illusion-bubble}, named \emph{jet-illusion bubble}, were done in axial symmetry with 46\,560 cells, while the simulation for the second bubble, named \emph{fast-jet bubble}, see Sec.~\ref{sec:fast-jet-bubble}, is done in full 3D with a mesh of about 6.5 million cells. In both cases, the computational mesh is composed such that it is finest in the core region of the bubble, where the cells are oriented in a Cartesian way, and radially coarsens outwards, with a polar orientation of the cells. The latter allows for an alignment of the cells with the bubble surface during a major part of the bubble evolution.
The overall mesh size is usually chosen 80 times the maximum bubble radius in the case of unbounded liquid or semi-unbounded liquid \citep{Koch-2016}. Both simulations are done for the semi-unbounded liquid, even though the cuvette for the jet-illusion-bubble of Sec.~\ref{sec:jet-illusion-bubble} was roughly 8 times smaller in the experiment. This, however, will not affect the validity of the results presented, as will be described later.
The axisymmetric case was set up with a spatial resolution of 1.5\mum\ and the 3D case with a resolution of 1.8\mum\ in the region of the bubble. In the axisymmetric case, the initial bubble shape is a cylinder of 9\mum\ radius and 116\mum\ height with the center at 216\mum\ above the solid boundary. In the 3D case, the initial bubble is a sphere with a radius of 20\mum\ with a distance of 21.8\mum\ to the solid boundary. In both cases the initial internal pressure is in the range of Megapascal (\emph{energy deposit case} \citep{Lauterborn-2018}), yielding a spherical equivalent maximum radius of $\approx 615\mum$ for the jet-illusion bubble and $\approx 495\mum$ for the fast-jet bubble. The initial velocity is zero in both cases. The adaptive time step is set to not exceed 8 times the acoustic Courant number and 0.2 times the fluid flow Courant number during maximum expansion and is limited to 1 times the acoustic Courant number during initial expansion and final collapse. Zero velocity and zero pressure gradient boundary conditions are applied to the solid boundary below the bubble, while the far liquid boundaries are kept transmissive for waves at ambient pressure $p_{\infty} = 101315\u{Pa}$ on average.

\section{Results for the static bubble}\label{sec:results_static}
\begin{figure}[htb]
     \begin{center}
     \includegraphics[width=0.6\tw]{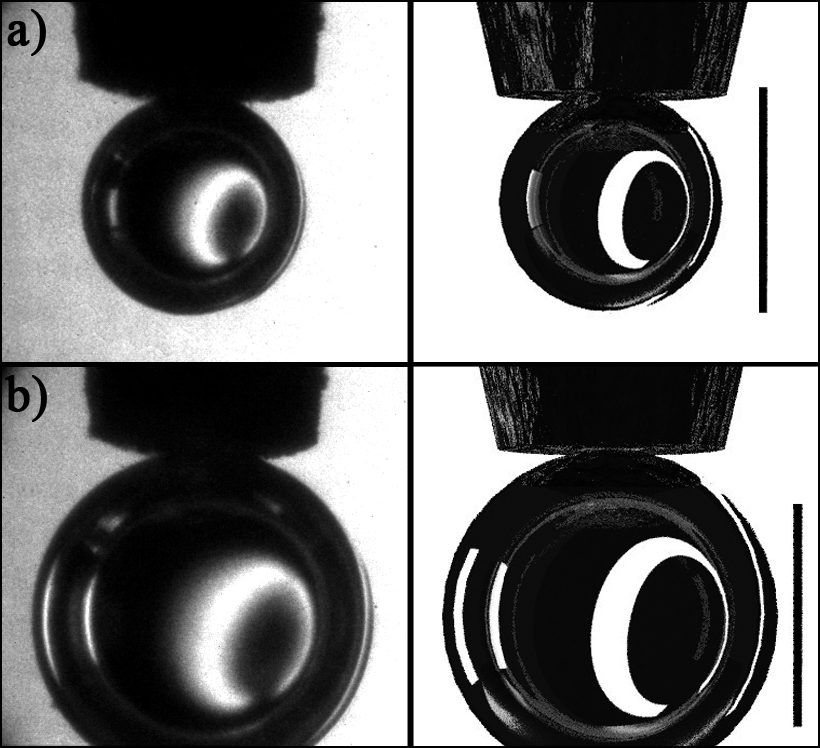}\\
     Experimental bubble\hspace{3cm} \blender-ray-traced bubble \\
     \end{center}
    \caption{Comparison of the experimentally obtained images (left) of a static, sub-millimeter bubble and their \blender-ray-traced representation (right). Top and bottom rows differ by the position of the side flash. The bar indicates a length of 250\mum.
    }
    \label{fig:staticbubbleresults}
\end{figure}

The image ray-traced from the perspective of Fig.~\ref{fig:setup_static_bubble_photo} is shown in Fig.\,\ref{fig:staticbubbleresults}. The upper row shows the outcome from the exact setup of Fig.\,\ref{fig:setup_static_bubble}, while the lower row depicts the outcome of the same setup but with the position of the side flash (Mecablitz) being rotated clockwise by 15$^{\circ}$ in the plane of the sketch with the bubble as rotation center. 

Clearly seen in each of the bubbles is the distorted image of the background ring flash tube, as well as the bright line on the outer right rim, caused by total reflection. Due to the asymmetry in the setup with respect to the line of sight this total reflection line is emphasized on the right side of the bubble. Also the influence of the position of the side flash is captured correctly by \blender: the rotated position of the flash enhances both refraction and total reflection.

\section{Results for the jet-illusion bubble}
\label{sec:jet-illusion-bubble} 
The \emph{jet-illusion bubble} serves as an example to show that refraction can be deceiving and apparently triplicate the jet velocity. The overlay-method can be used to infer the precise shape of a strongly aspherical bubble and correct the jet velocity value, even though the underlying simulation disregards 3D effects that might result from the small confinement of the cuvette.

\begin{figure}
    \centering
    \includegraphics[width=0.65\tw]{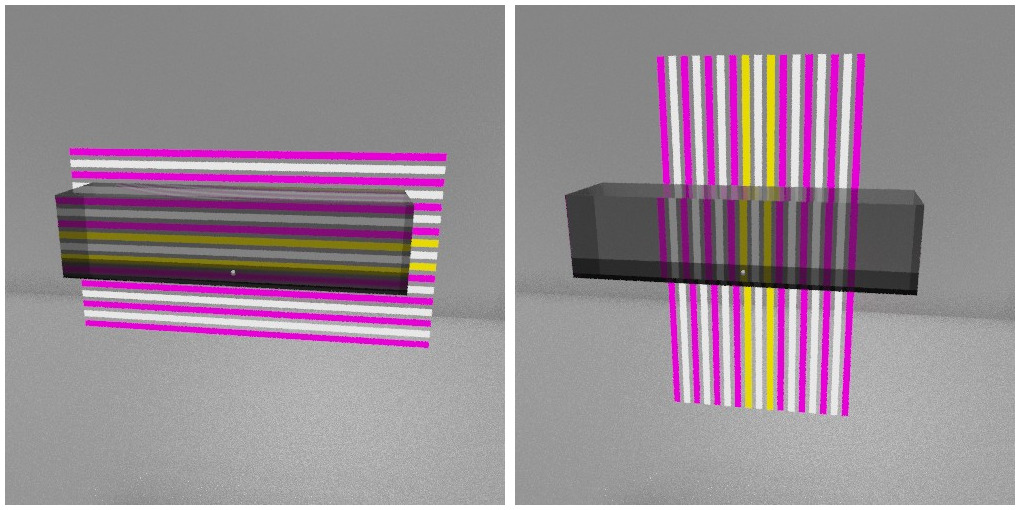}\\
    \includegraphics[width=\tw]{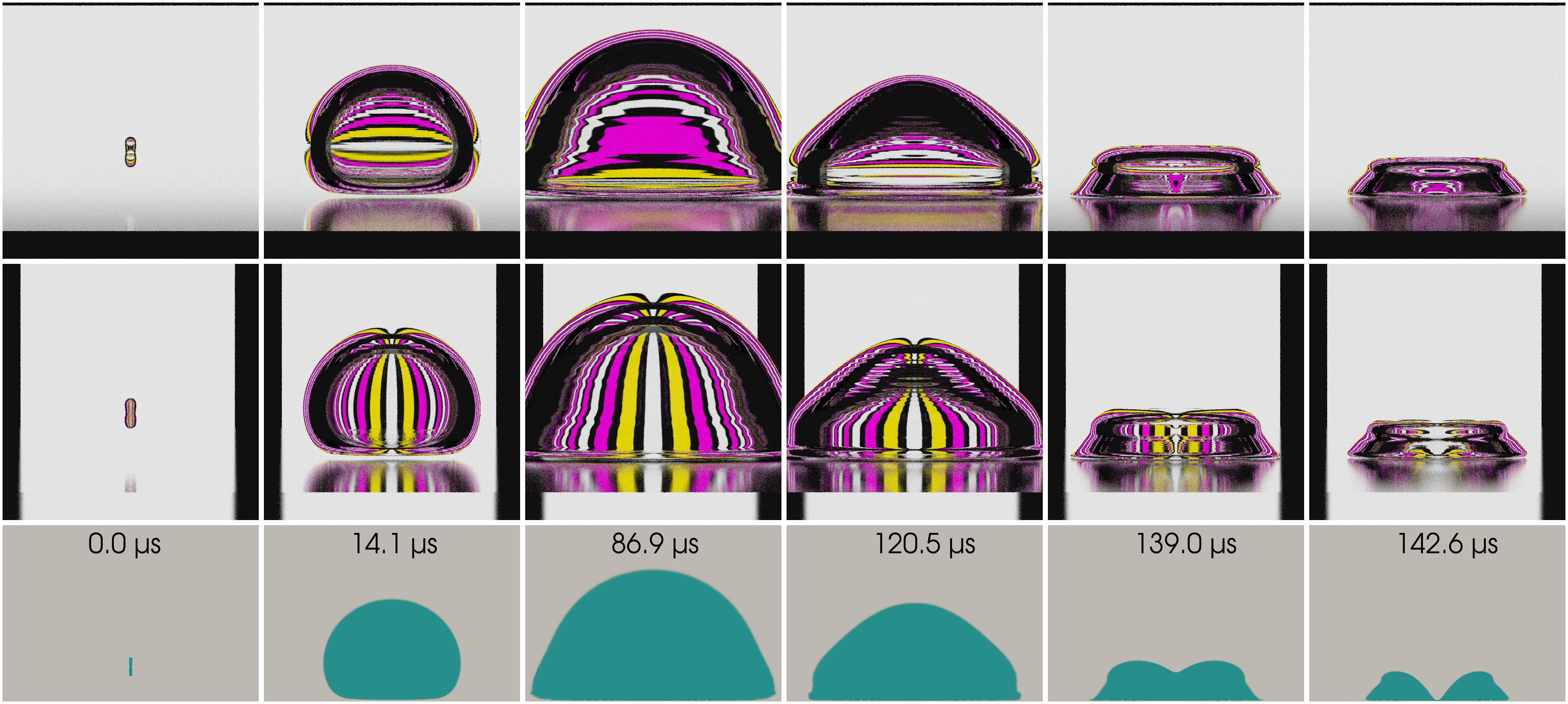}
    \caption{
     Ray-tracing of a regular, symmetric illumination pattern with alternating color (pink and white -- yellow next to the center) behind the small cuvette shown here with darker glass and lighter bubble for better contrast (first row). Note that the cuvette is rotated by 90$^\circ$ to the left compared to Fig.~\ref{fig:setup_dynamical_bubble}. Simulated bubble with illuminating rectangle in horizontal alignment (second row) and in vertical alignment (third row). The bubble is mirrored in the glass surface. Fourth row shows the bubble cross-section (not ray-traced). Frame width of the ray-traced bubbles is 1162\mum, the one of the cross-section is 1600\mum.} 
    \label{fig:pattern_illumination}
\end{figure}

As an example of the expected dynamics, distinct time steps of a simulation performed with the code described in Section~\ref{sec:code} are shown in Fig.~\ref{fig:pattern_illumination}. The laser-induced breakdown creates an elongated plasma which is modelled by an initial gas cylinder of high internal pressure at a distance of 216\mum\ to the wall, resulting into a dimensionless distance of $D^\ast = 0.332$ (Eq.~\ref{eq:dstar}). Afterwards the bubble expands and collapses including an involution of its surface because of the hindered water inflow from the boundary side. The cross-section of the bubble is shown in Fig.~\ref{fig:pattern_illumination}, together with the refraction of a pattern of 17 straight illumination stripes of alternating color (white, yellow and pink) behind the bubble. Each stripe has dimensions of $50.8\u{mm} \times 1\u{mm}$ in length and width. The stripe-to-stripe distance is 1.6\u{mm} and the pattern is put in a distance of 10\u{mm} behind the bubble, ergo 5\u{mm} behind the cuvette. The overall size of the pattern is $50.8\u{mm} \times 26.8\u{mm}$. In one of the plots the pattern is put in vertical, in the other frame in horizontal orientation. The vertical and horizontal pattern alignment is ray-traced and compared to the cross-section of the bubble.
Illumination grids or patterns are used for example to correct lens aberration \citep{nobach}.
The pattern deformation gives insight into the refraction distortion. In contrast to lenses, the distortion here is beyond linear approximations. It can be seen that for the time of interest, i.e., the jetting phase around $139\mus$ to $142.6\mus$, it is challenging to get an illumination configuration where the jet and bubble interior are clearly separated.
That is why the illumination device has to be designed and placed accurately. 
The circular geometry of the flash tube in the experiment fulfilled this requirement by chance.

With the \textit{Imacon 468} camera the involution (liquid jet) was recorded. One can obtain 8 images only, but at very high frame rate (down to 10\u{ns} inter-frame time). Hence, the higher the frame rate, the better the triggering circuit must be. Here the authors show some of the best results obtained by manually varying the delay time until the camera trigger happens to exactly match the time of jet formation during the bubble collapse phase. 

\begin{figure}[htb]
    \centering
    \includegraphics[width=0.9\tw]{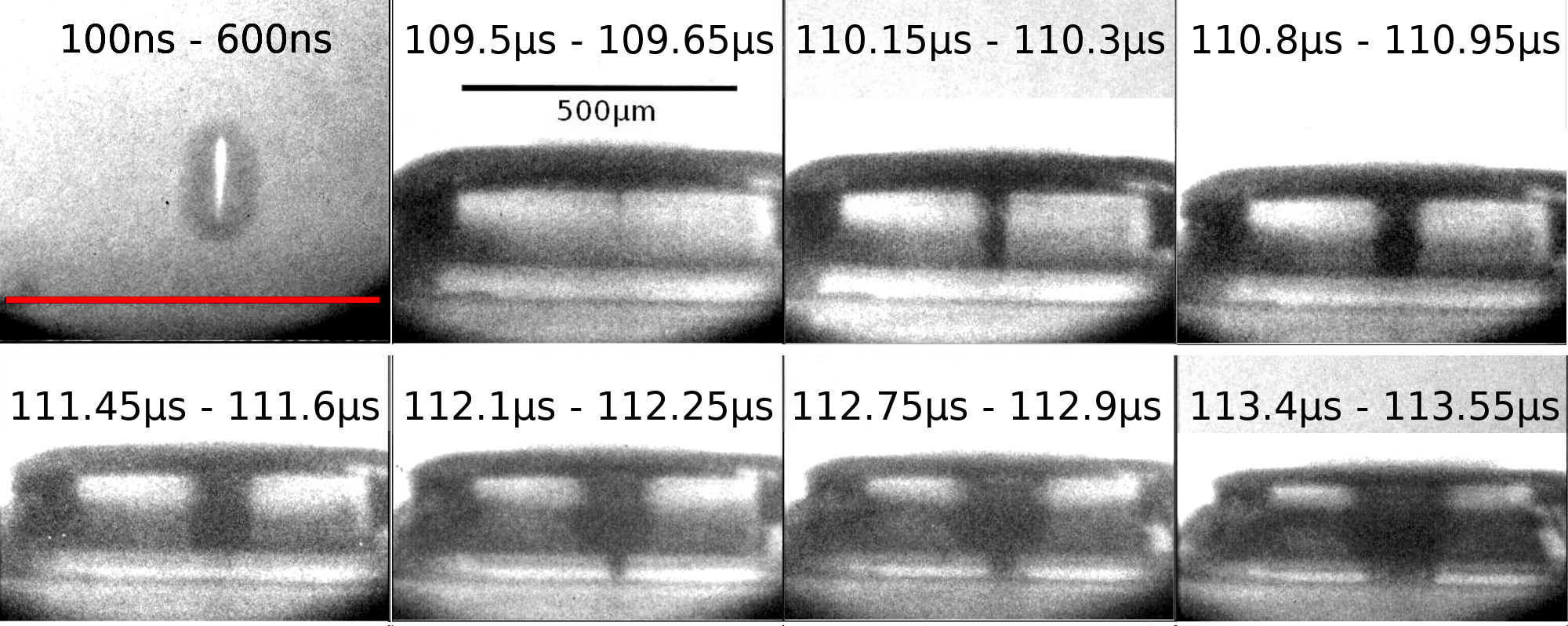}
    \caption{Liquid jet piercing the bubble, captured at 1.538\u{Mfps}. The first frame shows she plasma of the laser-induced breakdown. The red line in the first frame indicates where the solid boundary (glass) is located in the frames. 
    }
    \label{fig:THE_JET}
\end{figure}

The results in Fig.\,\ref{fig:THE_JET} were obtained with 150\u{ns} exposure time and an inter-frame time of 500\u{ns} between the end of the previous frame and start of the next frame. The glass surface wall is located at the bottom of the frames, indicated by the red line in the first frame, so the image is rotated by 90$^\circ$ compared to the setup sketch in Fig.\,\ref{fig:setup_dynamical_bubble}.
The first frame shows the plasma formation at the beginning. The times and spatial scale are indicated within the frames. The jet formation is seen clearly by a dark shadow piercing the bubble from top to bottom and widening over time. There seem to be two windows of the outer bubble interface where the jet is seen through. One big window in the middle and a narrower at the bottom. In the latter, the jet only is visible from frame 6 ($t = 112.1 \mus\ $) onward, although it has pierced the bigger window already in frame 3. 
To conclude, there is some nonlinear distortion of the apparent jet speed by refraction, since it is known that the velocity of the liquid jet is nearly perfectly constant over time and homogeneous in space \citep{Lauterborn-2018}. 
If the experiment was evaluated only by pixel counting, the following minimum jet speed would be found, taking the time between the end of the third frame and the end of the second frame of the top row:
$$ \frac{500\mum/352\u{px}}{110.3\mus - 109.65\mus}\cdot 80\u{px} \approx 175\u{m/s} $$

\begin{figure}[htb]
\centering\vspace*{-5ex}
\includegraphics[width=0.6\tw]{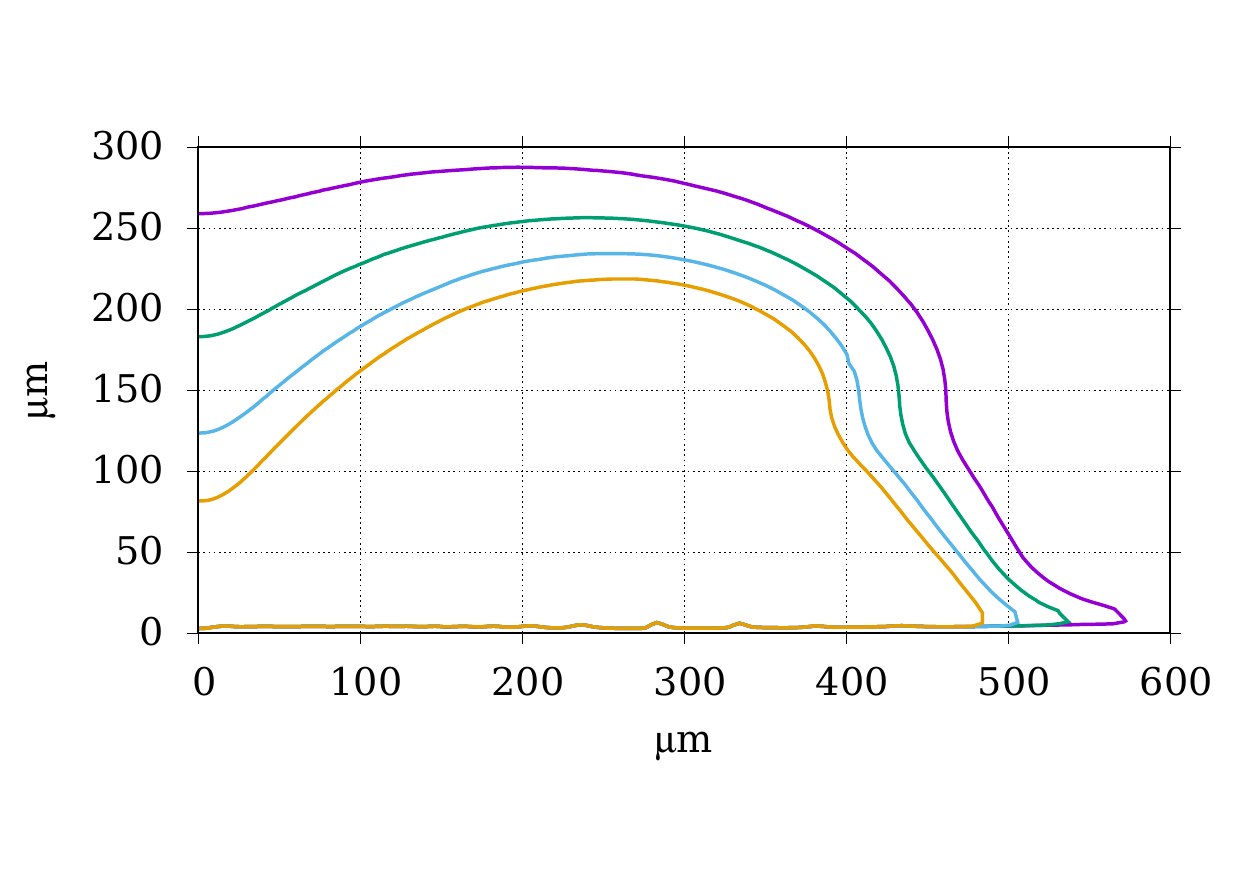}
\vspace*{-6ex}
\caption{The contour lines of half of the cross-section of the numerical bubble in Fig.~\ref{fig:pattern_illumination} for several instants of time during jetting. Axis of symmetry is located on the left side.}
\label{fig:involution_dynamics_simu}
\end{figure}

\begin{figure}[htb]
\centering
\includegraphics[width=0.99\tw]{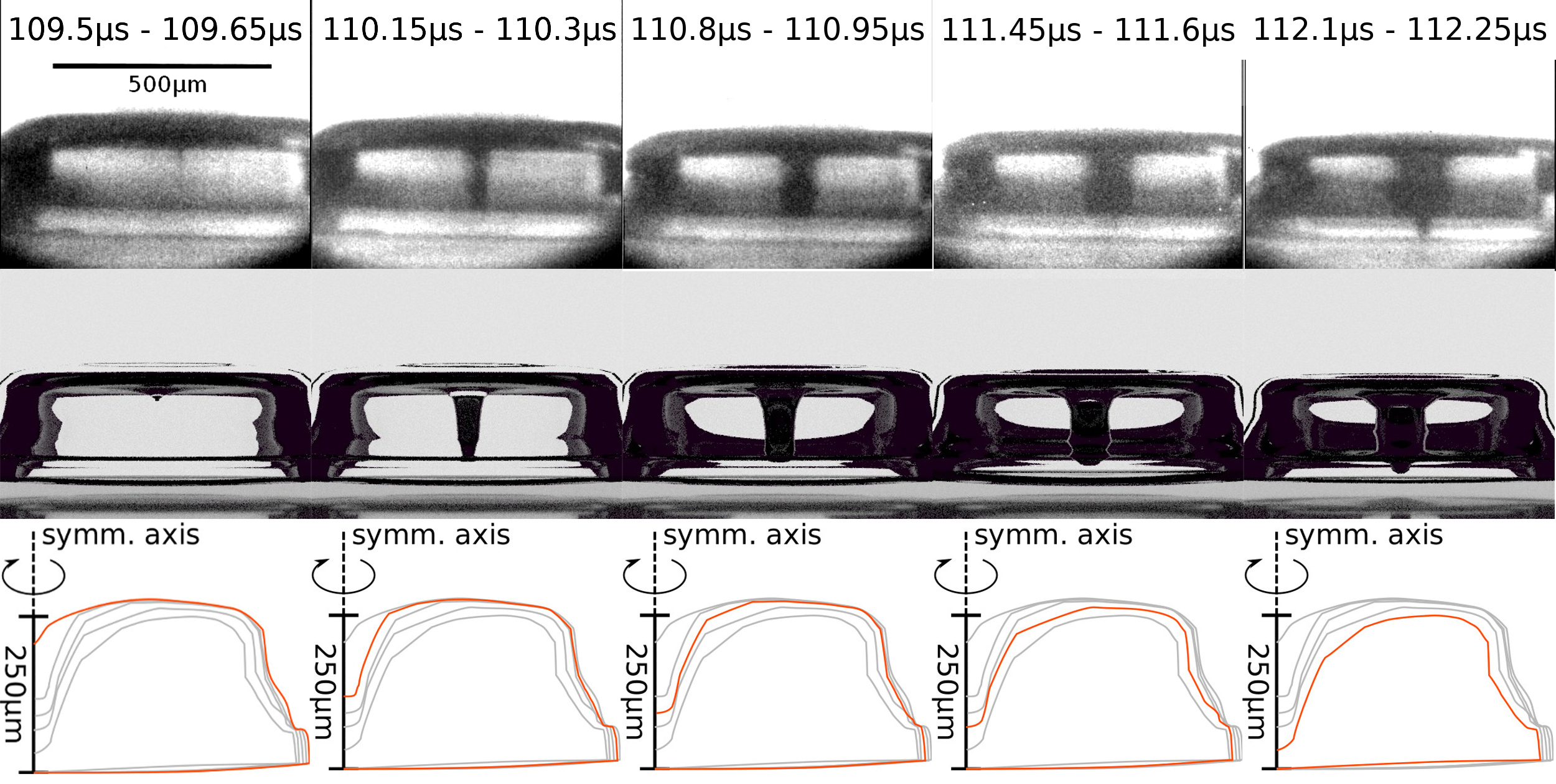}
\caption{The \textit{jet-speed illusion}. Top row: experimental photographs. Second row: ray-tracing of the manually adjusted profile that is defined by the curve shown in the third row.
}
\label{fig:thejetfusion}
\end{figure}

The appearance of light and shadow features highly depends on slightest changes of the interface curvature due to the nonlinearity in Snell's law. For determining the correct bubble shape from the photographs in Fig.~\ref{fig:THE_JET}, the bubble contour from the numerical simulation in Fig.~\ref{fig:pattern_illumination} is taken as a basis. The simulation gives a bubble involution dynamics as shown in Fig.~\ref{fig:involution_dynamics_simu}. 
The contour now is imported into \blender\ and reduced in grid complexity while keeping the shape. It is then rotated and extruded around the axis of symmetry. The curvature of the contour profile curve determines the curvature of the resulting object, the \textit{bubble probe}. The shape of the bubble probe can be adopted such that it resembles the experimental one to high level of detail by manually altering parts of the profile curve while watching the ray-tracing image outcome. This process has been applied to the bubble contour from Fig.~\ref{fig:involution_dynamics_simu} and is shown in Fig.~\ref{fig:thejetfusion}. 
\begin{figure}
    \centering
    \includegraphics[width=0.33\tw]{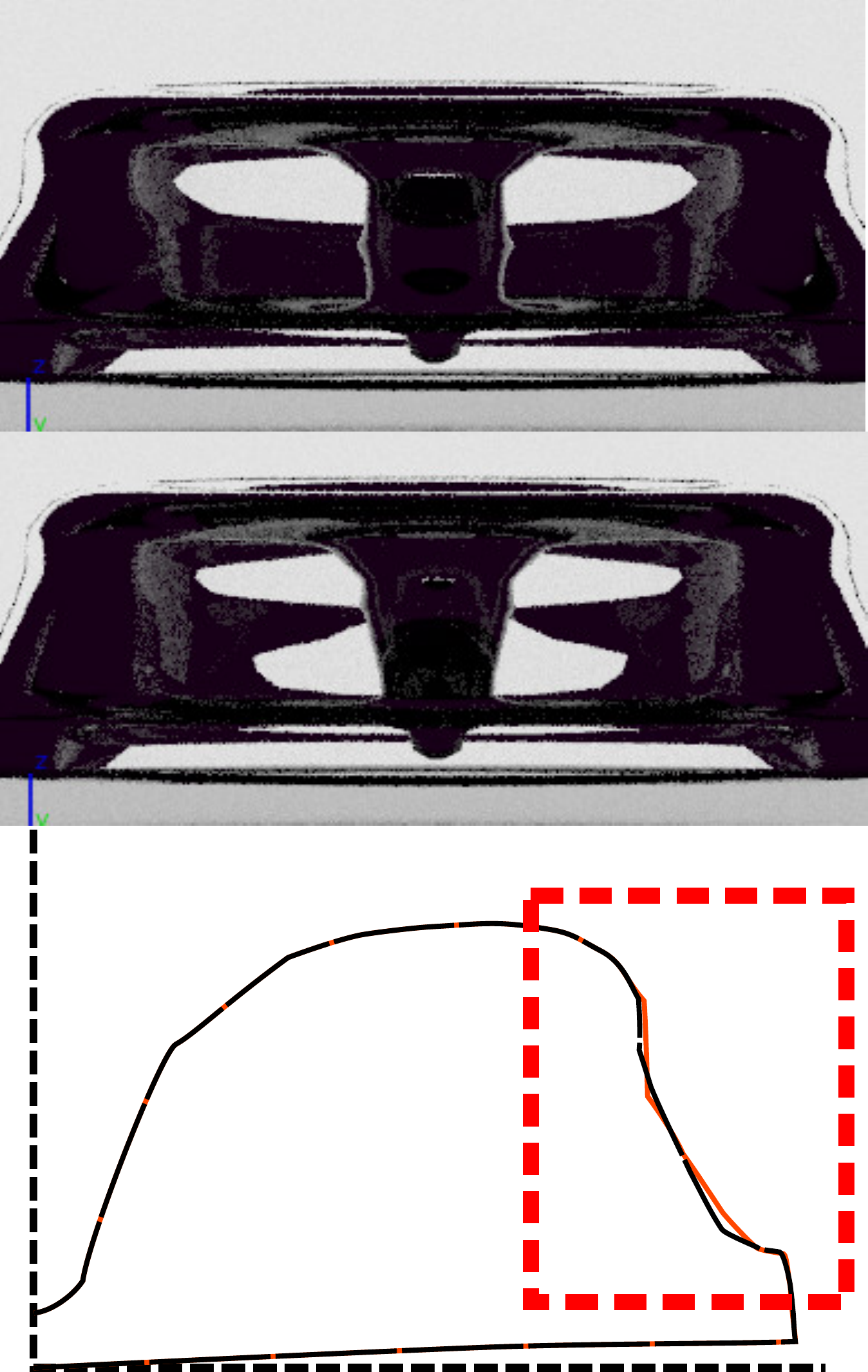}
    \caption{Example of the effect of a slight change of the bubble interface shape. The black contour line produces the second ray-tracing frame, whereas the original orange line produces the first frame. 
    }
    \label{fig:precision}
\end{figure}

Taking all light and shadow features into account, the bubble shape to obtain highest correlation becomes quite definite.
In some parts of the profile curve even a change in inclination by only a fraction of a degree results, e.g., into a fully light or half dark bubble interior. An example of the sensitivity of the shape is given in Fig.~\ref{fig:precision}, where the effect of a slight change in the interface curvature changes the image.

Now the jet velocity can be recalculated using the bubble contours in the bottom row of Fig.~\ref{fig:thejetfusion}:
$$ \frac{250\mum/246\u{px}}{110.3\mus - 109.65\mus}\cdot 87\u{px} \approx 57\u{m/s}, $$
giving approximately one third of the value that was derived from the photographs.

It is the first time known to the authors that the shape of a jetting cavitation bubble was inferred from the experiment with such high precision. With the successful application of the overlay-method, many more possibilities arise. One could think of, e.g., a specialized setup with an illumination grid in order to deduce the bubble shape automatically via an algorithm. Furthermore, it opens up the possibility to validate CFD codes to an unprecedented level of detail.

\section{Results for the fast-jet bubble}\label{sec:fast-jet-bubble}
\begin{figure}[htb]
    \centering
    \includegraphics[width=\tw]{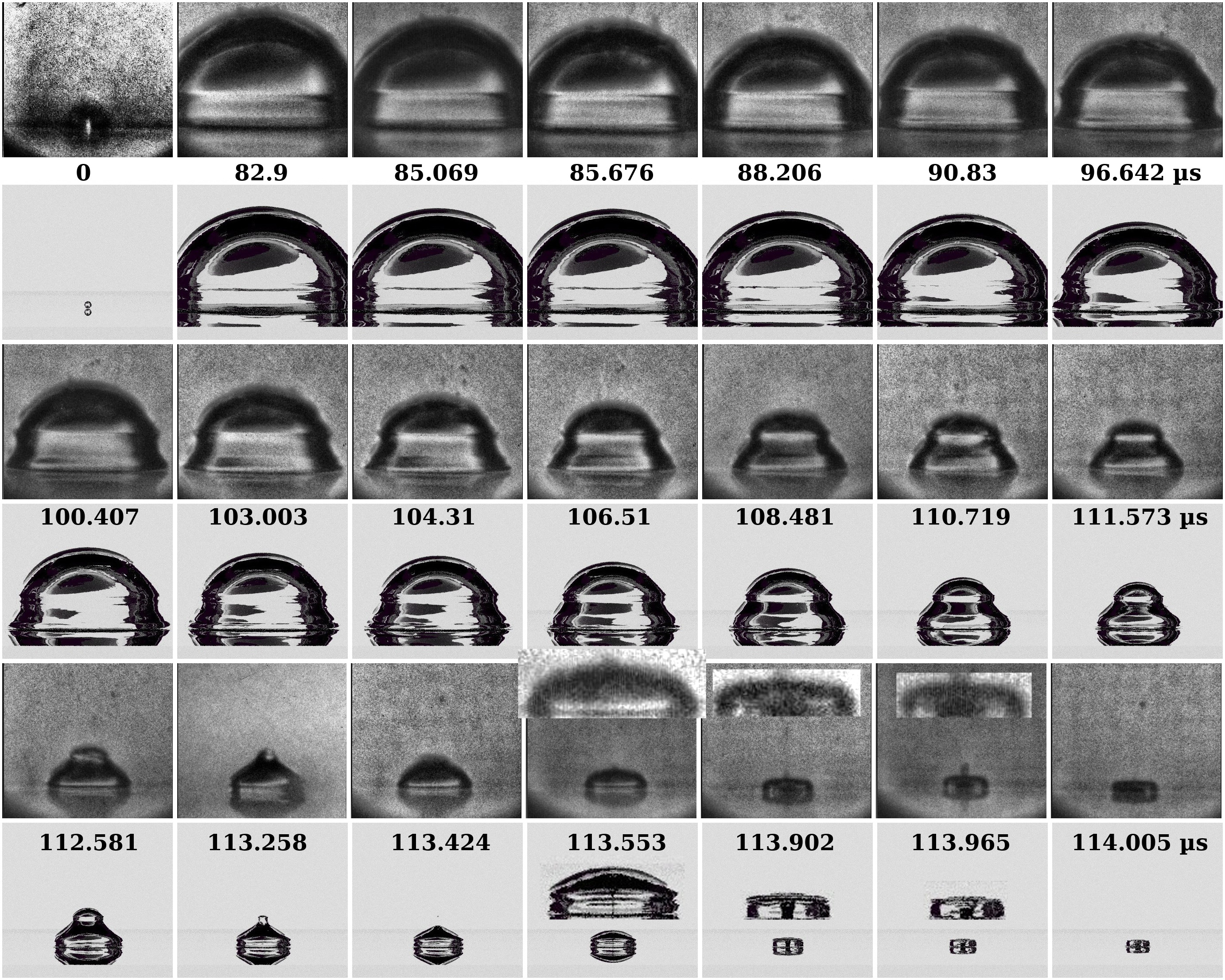}
    \caption{
    Bubble generated directly at the solid boundary. Odd rows are experimental recordings, even rows are ray-tracing images of the numerical simulation. Exposure time of the experimental frames is 150\,ns, except for the first frame with the plasma, which has an exposure time of 500\,ns and is enhanced in contrast. 
    The \blender-ray-tracing of the full 3D simulation of Fig.~\ref{fig:septembersequence_simu} at instants of time resembling the ones in the experiment is shown in the even rows. The width of the frames is 664.5\mum. For the three important frames for the jet, the bubble is also magnified and enhanced in contrast. At 113.553\,\mus\ the fast, needle-like jet can best be detected. 
    }
    \label{fig:septembersequence}
\end{figure}

\begin{figure}[htb]
    \centering
    \raisebox{80pt}{\textbf{a)}}\includegraphics[width=0.45\tw]{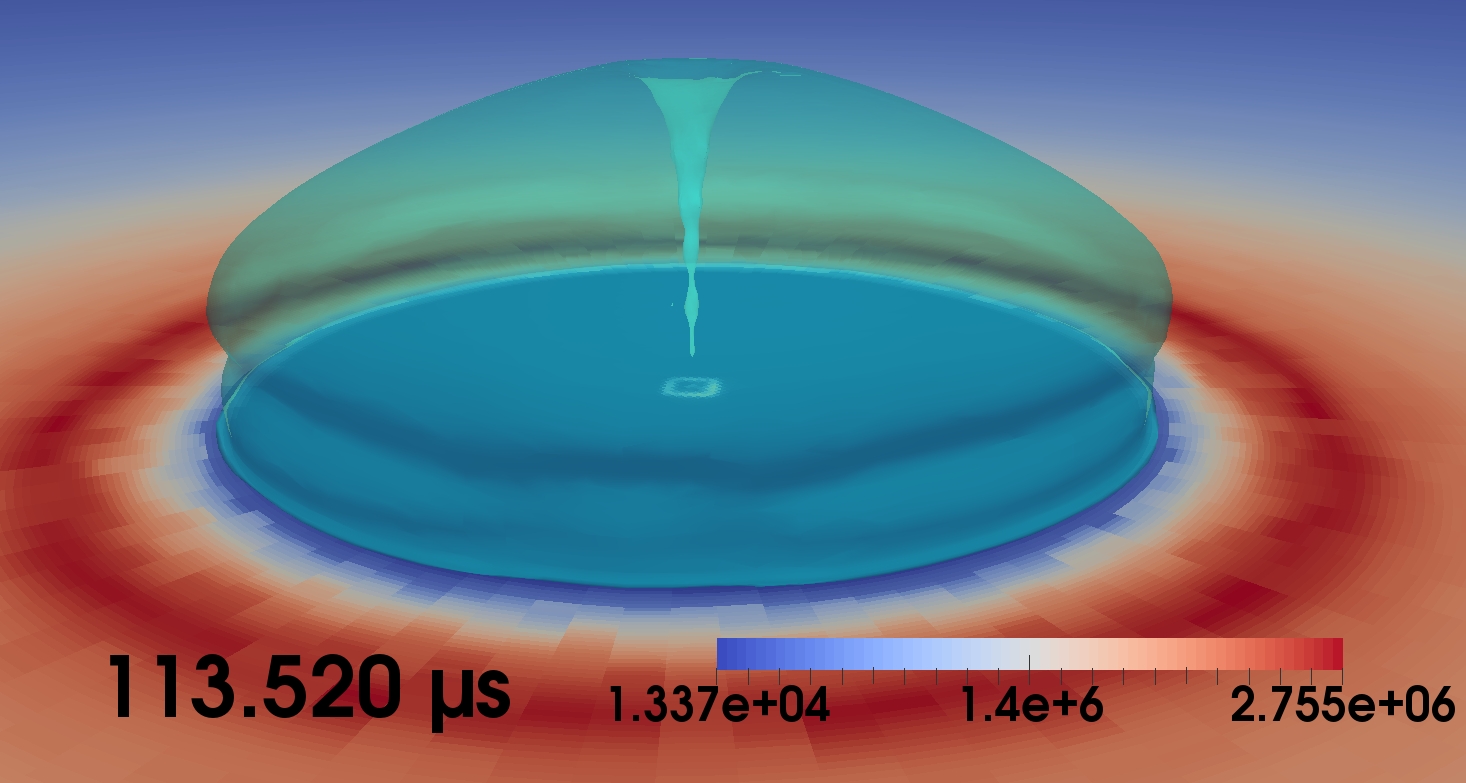}
    \raisebox{80pt}{\textbf{b)}}\includegraphics[width=0.45\tw]{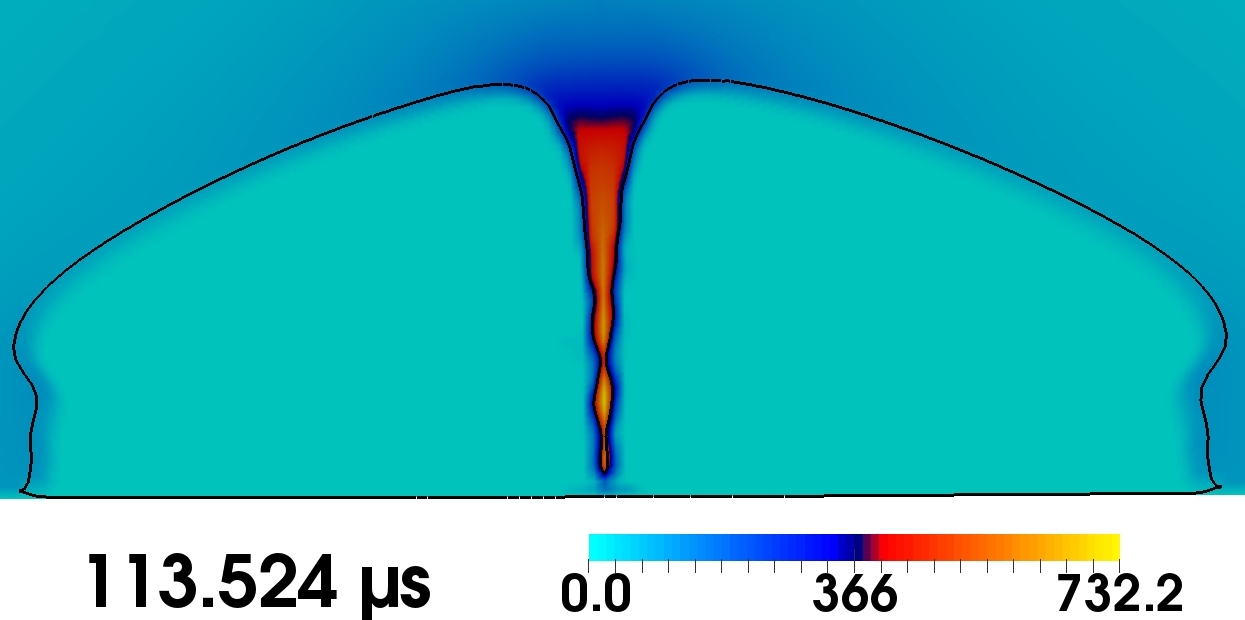}
    \caption{Underlying full 3D simulation of the fast-jet bubble. a): contour plot of the bubble interface with the pressure in Pascal plotted in the plane of the solid boundary. b): Cross-section through the same bubble, plotted with the liquid velocity in m/s. 
    }
    \label{fig:septembersequence_simu}
\end{figure}

The confidence of the ray-tracing engine gained by the excellent agreement for the static bubble and the first dynamical bubble led to the application for a bubble with the hardly measurable phenomenon of the fast-jet. Actually, it were the difficulties with the experimental fast-jet that started the overlay-method for better extraction of results from experiments. The \emph{fast-jet bubble} was recorded with high temporal precision and generated at a dimensionless distance of $D^\ast = 0.04$. It is taken as an example where the overlay-method may give an indirect hint for the existence of hardly measurable phenomena.
The result is shown in Fig.~\ref{fig:septembersequence}. The odd rows show the photographs of the experiment and the even rows show the ray-tracing of the 3D CFD simulation of the bubble (depicted representatively in Fig.~\ref{fig:septembersequence_simu}). The frames again are rotated such that the glass surface boundary is located in the lower part. The first frame shows the plasma of the laser-induced breakdown, and from the second frame onward the collapse phase of the bubble is shown. 
The sequence is a montage of 7 measurements in total. The exposure time was the same for all measurements (150\,ns) while the inter-frame time varied from 1\mus\ to 350\,ns. By stacking the single frames of the reproducible measurements together, a time resolution down to $40\u{ns} \stackrel{\wedge}{=} 25\u{Mfps}$ could be achieved.
Here, the times of the simulation are taken to indicate the time progression.

The typical bell shape is seen, which was already reported in \citet{Benjamin-1966} and predicted in \citet{Lechner-2019} by means of numerical simulations in axial symmetry. The ray-tracing shows most light and shadow features of the experiment. The bubble is magnified in inlet frames in the last time steps in Fig.~\ref{fig:septembersequence}, so that it can be seen that the bubble was indeed pierced. 

A visualization of the numerical simulation at time $t= 113.52 \mus $ is given in Fig.~\ref{fig:septembersequence_simu}. The left frame shows the 3D bubble contour together with the pressure field at the solid surface. The right frame shows the velocity field in the liquid at a cross sectional plane through the bubble. With a spatial resolution of 1.8 \mum\ the numerical simulation predicts the formation of a fast jet with a speed of 736.6\, m/s. 

Figs.~\ref{fig:septembersequence} and \ref{fig:septembersequence_simu} together prove that a) the bubble shape including the fast jet is not an artefact of axial symmetry, but a full 3D feature and b) the bubble shape is well captured by the CFD simulations.
Taking these arguments together, it is most likely that there was a fast jet occurring also in the real experiment. Three main reasons are considered why it is not seen in the frames at times 113.424\mus\ and 113.553\mus: First, the jet width is still below resolution, because the resolution is lowered by the noise 
of the camera. Second, the focal plane of the interior of the bubble is different to the silhouette and the depth of the focal plane was not deep. Third, there is still some possibility that the illumination geometry does not provide jet and bubble interior separation.
This shows that the overlay-method multiplies scientific interpretation possibilities for (two-phase) fluid flows, even reaching out to hardly measurable phenomena.

\section{Conclusion}\label{sec:Conclusion}
By illuminating the results of the finite-volume simulations for single cavitation bubbles via ray-tracing, it was made possible to directly compare them to experiments with excellent agreement. Thus, it has been shown that the 3D editing and ray-tracing software \blender\ can be used as a scientific post- and pre-processing method. After validating the ray-tracing method with a static bubble attached to a needle, the method was applied to scientifically crucial experiments: Jetting bubbles in close vicinity of a solid boundary. The ray-traced images of the simulations were compared to experiments where bubble jets were recorded with ultra-high-speed imaging. Because of the congruence, it can first of all be deduced that the ray-tracing engine works well and the CFD simulation code produces valid results to a high level of detail for the tested scenarios. Assuming that the congruence allows for inverting the logical chain, the following criteria relevant for the physics of single cavitation bubbles could be deduced:
\begin{itemize}
    \item For small initial distances of bubble generation to the solid boundary there might be jets seen in the experiments that optically appear much faster than they actually are. This fact arises from the interface shape alienation from a simple sphere and thus complications in refraction patterns.
    \item For an initial bubble distance tending to zero ($D^\ast =0$), probably there exists a fast collapse jet with a velocity exceeding 700\,m/s.
    An exact number cannot be given here, since the time consuming 3D simulation was only performed for one resolution, lacking a convergence study. The fast jet was predicted in \citet{Lechner-2019} via calculations in axial symmetry only. Here it is shown that this jet also is formed in a full 3D simulation and the results of the ray-traced bubble shape compare well to the ones obtained in the experiment.
\end{itemize}

In general, it was shown that the ray-tracing overlay-method improves interpretation possibilities of experimental results. The method indirectly confirmed the validity of the code used to simulate the cavitation bubbles.

\begin{journal}
  \printbibliography
\end{journal}
\begin{arXiv}
   \bibliography{MK_raytracing_2020}
\end{arXiv}

\section{Declarations}
\subsection{Funding} 
The work was funded by the German Research Foundation (Deutsche Forschungsgemeinschaft, DFG) under the project Me 1645/8-1.
J. M. Rossell\'o acknowledges support by the Alexander von Humboldt Foundation (Germany) through a Georg Forster Research Fellowship.
\subsection{Conflicts of interest/ Competing interests}
The authors declare that they have no conflict of interest.
\subsection{Availability of data and material}
Not applicable.
\subsection{Code availability}\label{sec:codeavail}
The code to produce Fig.~\ref{fig:Fig1}b is published and maintained at \citet{python-raytracer}. \blender\ is available at \texttt{blender.org}. The version used was 2.79b. The open source CFD package \texttt{foam-extend-4.0} is available at \citet{fext40}. An altered version of the solver \texttt{compressibleInterFoam} was used, as described in \citet{Koch-2016}.

\end{document}